\documentclass[aps,twocolumn,prd,superscriptaddress,letterpaper]{revtex4-1}
\usepackage{dcolumn}

\usepackage[pdftex]{graphicx}
\usepackage[usenames,dvipsnames]{color}     
\usepackage{verbatim}
\usepackage{url}
\usepackage[latin1]{inputenc}
\usepackage[T1]{fontenc}
\usepackage{amsmath, amssymb}
\usepackage{braket}
\usepackage{subcaption}
\usepackage{xspace}
\usepackage{ftnxtra}
\usepackage{fnpos}
\usepackage{mathtools}
\usepackage{caption}
\usepackage{finesse}
\usepackage{lineno}
\usepackage{fancyvrb}
\DefineVerbatimEnvironment{finesse}{Verbatim}
{formatcom=\small}

\usepackage[pdftex,a4paper,pagebackref=true,pdfpagelabels=true]{hyperref}
\definecolor{linkcolor}{rgb}{.8,0,0}
\definecolor{urlcolor}{rgb}{0,0,.7}
\definecolor{citecolor}{rgb}{0,.5,0}
\definecolor{acrocolor}{rgb}{0,0,.7}
\hypersetup{bookmarksopen,colorlinks=true}
\hypersetup{pdfstartview=FitH}
\hypersetup{linktocpage=true,bookmarksnumbered=true}
\hypersetup{plainpages=false,breaklinks=true}
\hypersetup{linkcolor=linkcolor,citecolor=citecolor,urlcolor=urlcolor}

\begin{document}




 \title{Multi-spatial-mode effects in squeezed-light-enhanced interferometric gravitational wave detectors}

\author{Daniel T\"oyr\"a}
\email{dtoyra@star.sr.bham.ac.uk}
\affiliation{School of Physics and Astronomy and Institute of Gravitational Wave Astronomy, University of Birmingham, Birmingham B15 2TT, United Kingdom}
\author{Daniel D. Brown}
\affiliation{School of Physics and Astronomy and Institute of Gravitational Wave Astronomy, University of Birmingham, Birmingham B15 2TT, United Kingdom}
\author{McKenna Davis}
\affiliation{Rhodes College, Memphis, TN 38112, USA}
\author{Shicong Song}
\author{Alex Wormald}
\affiliation{School of Physics and Astronomy and Institute of Gravitational Wave Astronomy, University of Birmingham, Birmingham B15 2TT, United Kingdom}
\author{Jan Harms}
\affiliation{Universit\`a degli Studi di Urbino "Carlo Bo", I-61029 Urbino, Italy}
\author{Haixing Miao}
\author{Andreas Freise}
\email{adf@star.sr.bahm.ac.uk}
\affiliation{School of Physics and Astronomy and Institute of Gravitational Wave Astronomy, University of Birmingham, Birmingham B15 2TT, United Kingdom}
\date{\today}

\begin{abstract}

Proposed near-future upgrades of the current advanced interferometric gravitational wave detectors include the usage of frequency dependent squeezed light to reduce the current sensitivity-limiting quantum noise. 
We quantify and describe the degradation effects that spatial mode-mismatches between optical resonators have on the squeezed field. These mode-mismatches can to first order be described by scattering of light into second-order Gaussian modes. 
As a demonstration of principle, we also show that squeezing the second-order Hermite-Gaussian modes $\mathrm{HG}_{02}$ and $\mathrm{HG}_{20}$, in addition to the fundamental mode, has the potential to increase the robustness to spatial mode-mismatches. 
This scheme, however, requires independently optimized squeeze angles for each squeezed spatial mode, which would be challenging to realise in practise.
\end{abstract}

\maketitle

\section{Introduction}
\label{sec:intro}

\begin{figure}[t]
\begin{center}
\includegraphics[width=0.45\textwidth]{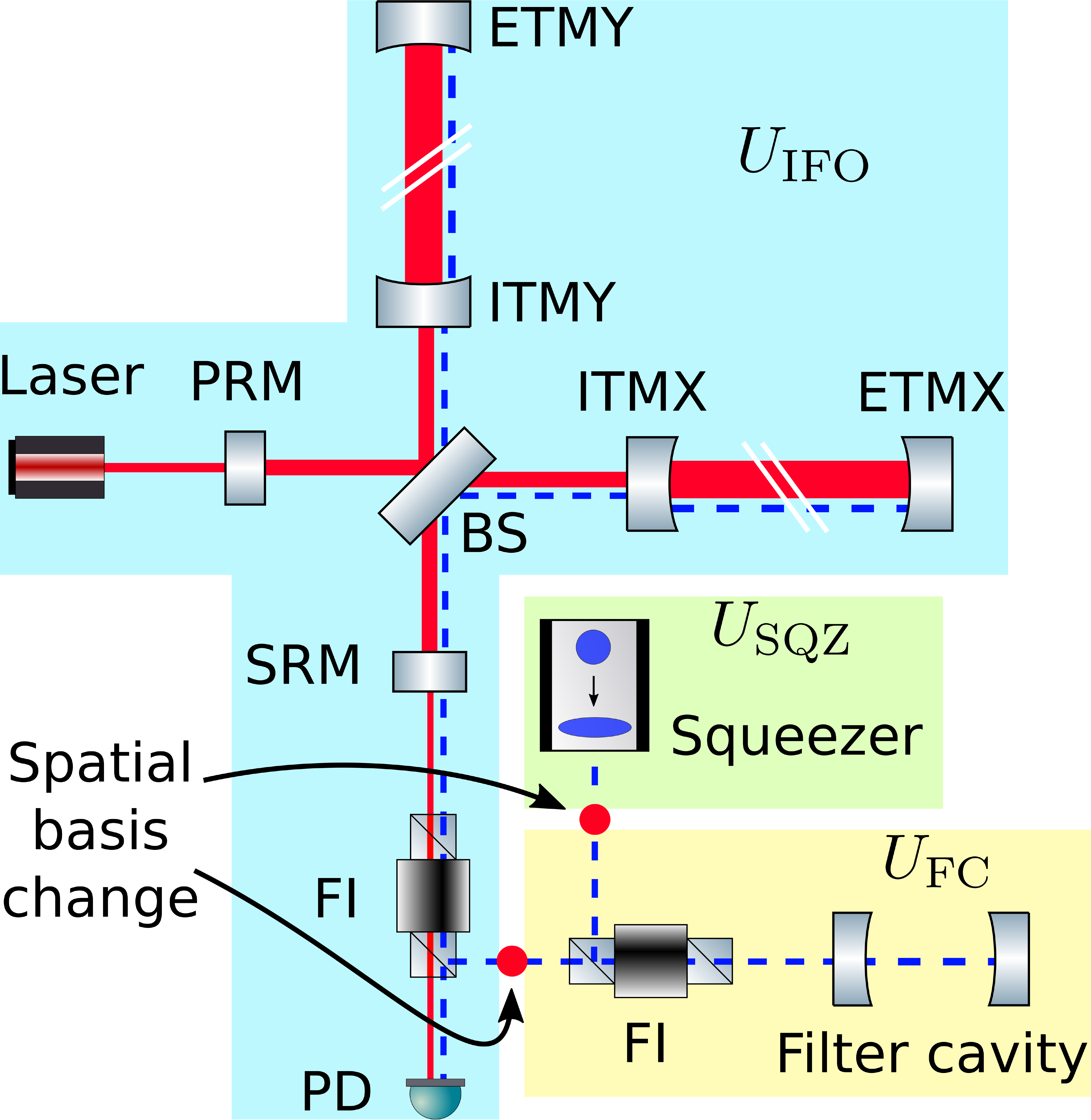}
\caption{The field emitted by the squeezer is reflected off a filter cavity to produce frequency dependent squeezed states. These states are injected into the interferometer through the signal recycling mirror. There are three potentially different spatial eigenbases in this setup: $U_\mathrm{FC}$ for the filter cavity (yellow background), $U_\mathrm{SQZ}$ for the squeezer (green), and $U_\mathrm{IFO}$ for the interferometer (blue), where the background colors indicate which basis that is used where. All the coherent laser power is in the fundamental mode of the interferometer basis.}
\label{fig:setup}
\end{center}
\end{figure}

The current advanced gravitational-wave detectors, e.g., the Advanced LIGO~\cite{AdvancedLIGO15} detectors, are dual-recycled Michelson interferometers with arm cavities, as shown in Fig.~\ref{fig:setup}. 
One of the limiting noise sources is quantum noise which arises from quantum fluctuations of light. 
To reduce the quantum noise over a broad-frequency band, one approach is to inject frequency dependent squeezed vacuum states into the dark port of the interferometer~\cite{Caves81, Unruh82}. 
These states are produced by the combination of a \emph{squeezer} and a \emph{filter cavity}, where the filter cavity generates the frequency dependency~\cite{Kimble02, Harms03, Chelkowski05}, such that the phase quadrature is squeezed for high frequencies and the amplitude quadrature is squeezed for low frequencies. 
This technology can be fitted into the current infrastructure~\cite{Evans2013, Kwee2014}, and is planned to be implemented in the next upgrade of the current observatories. \\

There are several practical imperfections that can influence the performance of this scheme, such as spatial mode-mismatches, optical losses, and phase noise~\cite{Yonezawa12, Kwee2014, Dwyer13}. This paper focuses on spatial mode-mismatches. Their effects on the squeezing can be categorized into two types. The first type is when a part of the squeezed states in the fundamental mode irreversibly scatters to higher-order modes, which has an effect similar to an optical loss. The second type is when the quantum states are allowed to coherently couple back and forth between the fundamental and higher-order modes. This type requires multiple interfaces where mode-mismatch induced scatterings occur. Particularly, there are two important such interfaces, located between the three components of interest in this work: the squeezer, the filter cavity, and the interferometer --- each to a good approximation having its own well-defined spatial mode basis. \\

Kwee \emph{et al.}~\cite{Kwee2014} studied the combined effect of these two types by considering mode-mismatches at the above mentioned interfaces. 
In this study, to better understand these two effects individually, we isolate them as much as possible by mode-mismatching one of the three components at a time, i.e., two components are always kept perfectly mode matched to each other. 
In contrast to Ref.~\cite{Kwee2014} and to what would be done in practice, the filter cavity is intentionally made to be resonant for higher-order modes within the frequency band of interest. 
On the one hand, this allows us to further study the interesting coherent scattering effect. On the other hand, it might also be relevant in reality for long filter cavities. \\

Additionally, we have looked into whether injecting \emph{multi-spatial-mode squeezing}, where two higher-order spatial modes are squeezed in addition to the fundamental mode, can provide robustness to mode-mismatches. The interesting spatial aspects of squeezed states have generated the relatively new field of quantum imaging~\cite{Kolobov89, Lugiato93, Lugiato02}, which has experimentally demonstrated the abilities of both generating squeezed higher-order Gaussian modes~\cite{Lassen06, Morizur11, Semmler16}, and combining different squeezed transverse modes~\cite{Treps03}. These are, in principle, the tools needed to produce the multi-spatial-mode squeezing considered in this paper. \\

\begin{figure}[t]
\begin{center}
\includegraphics[width=0.48\textwidth]{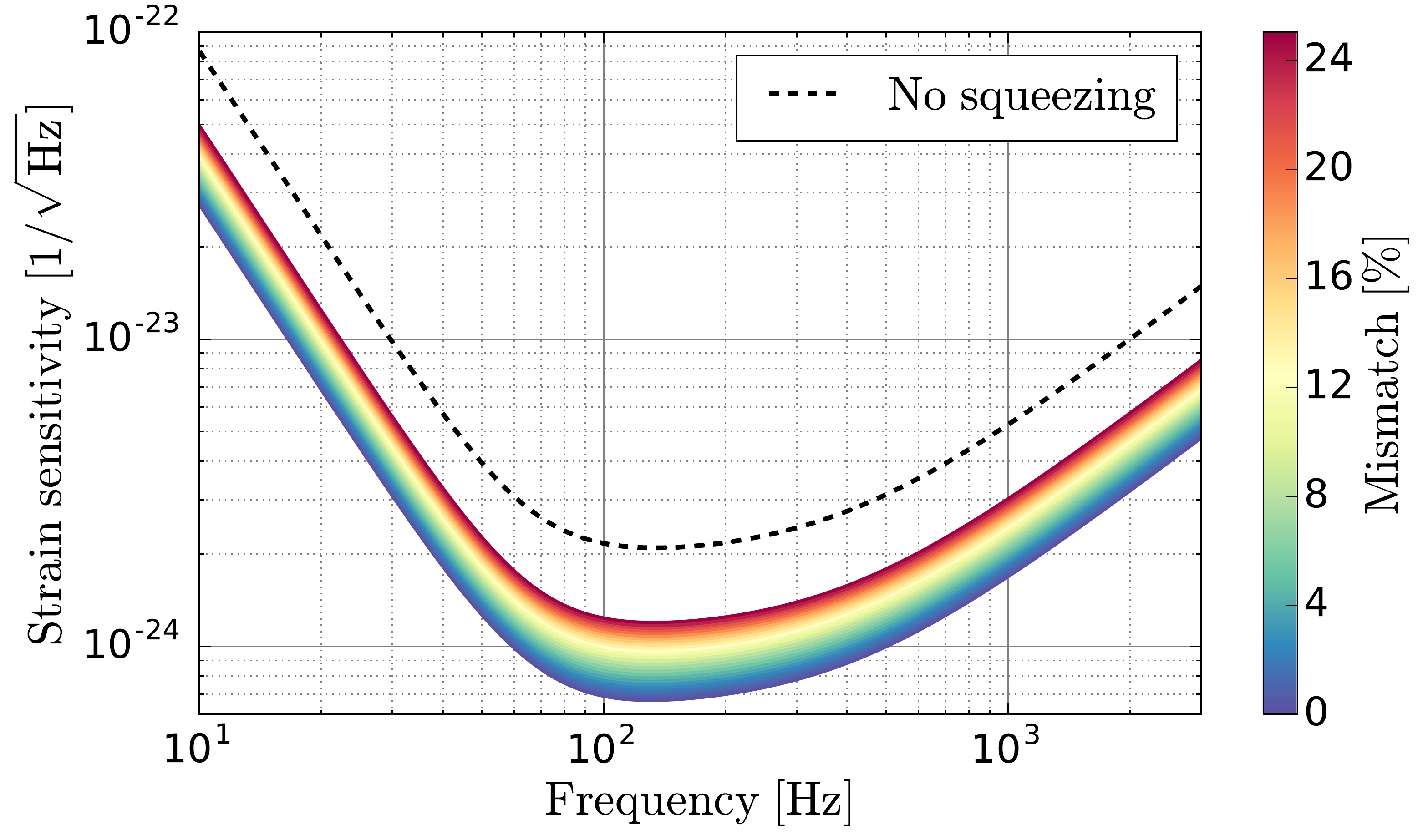}
\caption{The figure shows the quantum-noise-limited sensitivity for various levels of mode-mismatch between the interferometer and the filter cavity. The squeezer is kept mode matched to the filter cavity. This type of mode-mismatch creates a broad-frequency band squeezing degradation similar to an optical loss.}
\label{fig:ssm_ifo}
\end{center}
\end{figure}

\begin{figure}[t]
\begin{center}
\includegraphics[width=0.48\textwidth]{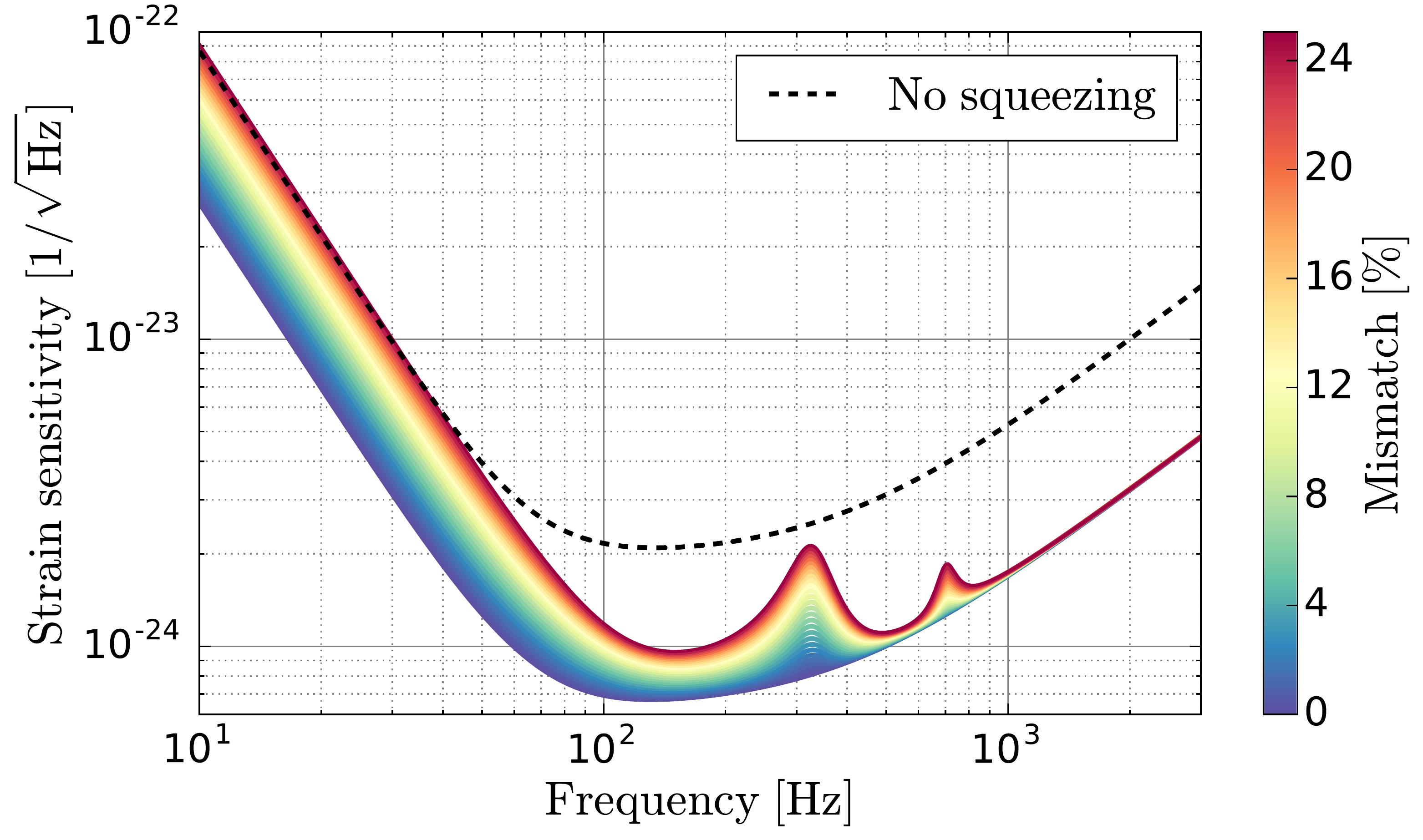}
\caption{In contrast to Fig.~\ref{fig:ssm_ifo}, the squeezer is kept mode matched to the interferometer. Around the resonance frequencies of the involved spatial modes, we experience squeezing degradation due to coherent mode-scattering.}
\label{fig:ssm_fc}
\end{center}
\end{figure} 

\begin{figure}[htb]
\begin{center}
\includegraphics[width=0.40\textwidth]{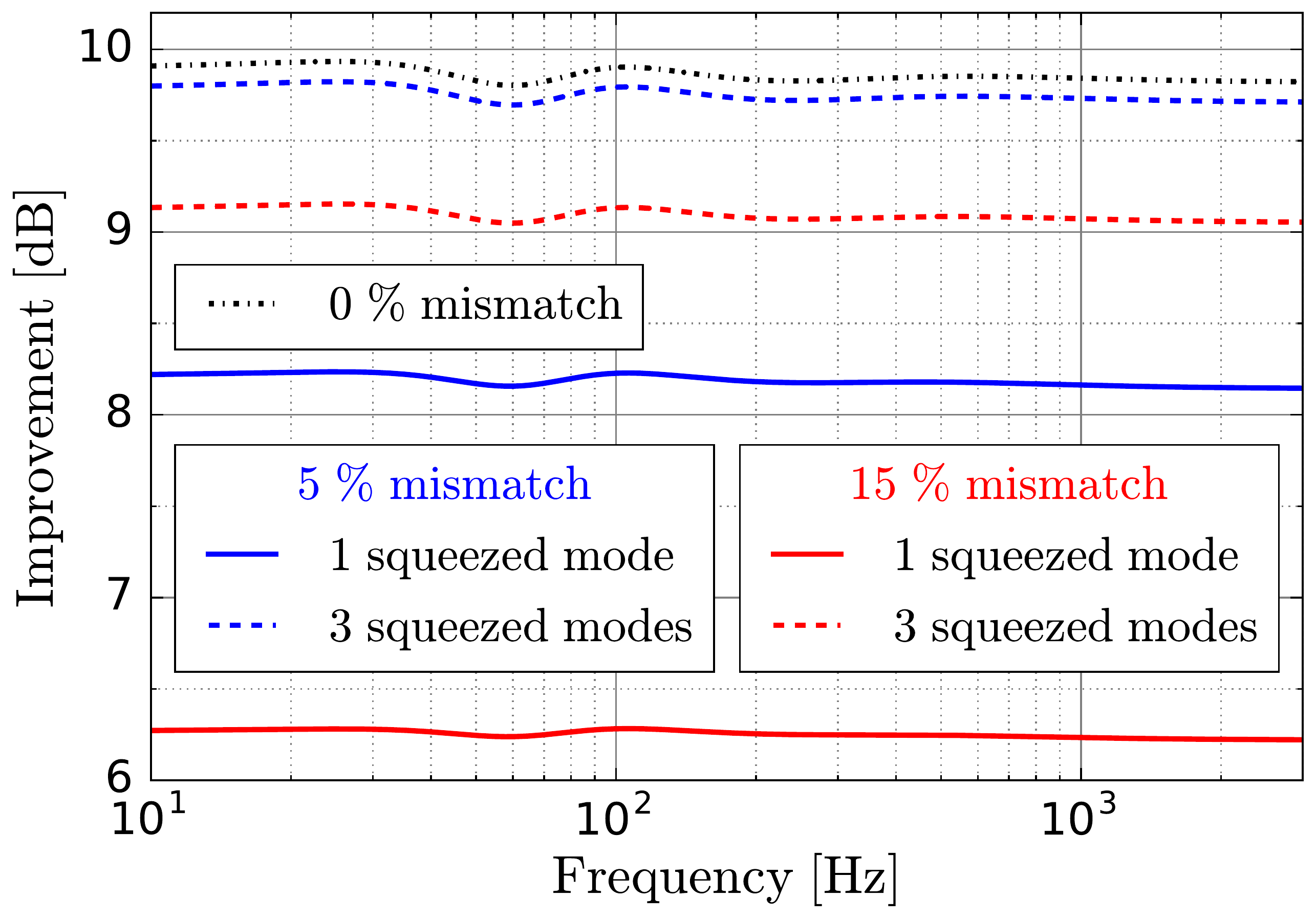}
\caption{The figure shows the improvement in dB that we obtain by squeezing the vacuum fluctuations that enters through the signal recycling cavity. The dashed traces indicates the improvement when squeezing 3 spatial modes, and the solid lines indicate the improvement when squeezing 1 spatial mode. This is shown for two different levels of mode-mismatch.}
\label{fig:msm_ifo}
\end{center}
\end{figure}

The key results of this paper are summarized as follows. 
In Fig.~\ref{fig:ssm_ifo} we show the quantum noise limited sensitivity for various levels of mode-mismatches between the interferometer and the filter cavity, while keeping the squeezer mode matched to the filter cavity. This mode-mismatch has the same effect as a lossy element between the filter cavity and the interferometer. The exact same effect is seen when mode-mismatching the squeezer to a mode matched filter cavity and interferometer. These results are consistent with the result obtained by Kwee \emph{et al.}~\cite{Kwee2014} in the high-frequency part of the spectrum. 

Figure~\ref{fig:ssm_fc} shows the result when the squeezer is kept mode matched to the interferometer instead of to the filter cavity. 
In this case, there are scattering points (spatial basis changes) before and after the filter cavity, which allows the squeezed states to coherently scatter to higher-order modes and then back to the fundamental mode. 
If a higher-order mode involved in this process picks up a different phase than the fundamental mode when reflected off the filter cavity, this mode-mismatch enables for potentially antisqueezed states to mix in with the squeezed states---which would be worse than just a loss. 
This coherent scattering effect can be seen in Fig.~\ref{fig:ssm_fc} at low frequencies where the fundamental mode is near-resonant while the higher-order modes are off resonance, and at the two local peaks where the second and fourth order modes are resonant while the fundamental mode is off resonance. 
These results are consistent with the low-frequency part of the spectrum obtained by Kwee \emph{et al.}~\cite{Kwee2014}. \\

Figure~\ref{fig:msm_ifo} shows the results obtained when letting the field emitted by the squeezer have squeezed states in the three Hermite-Gaussian modes $\mathrm{HG}_{00}$, $\mathrm{HG}_{02}$, and $\mathrm{HG}_{20}$. This is in contrast to above where only the $\mathrm{HG}_{00}$ mode was squeezed. Just as when generating Fig.~\ref{fig:ssm_ifo}, the filter cavity is mode-mismatched to the interferometer while the squeezer is kept mode matched to the filter cavity. The filter cavity is redesigned so that the second order modes have the same resonance condition as the fundamental mode, which is necessary to correctly rotate all the squeezed states. In addition, the squeeze angles of the second order modes have been independently optimized to maximize the broad-frequency band sensitivity. Figure~\ref{fig:msm_ifo} shows that, in principle, the injection of a multi-spatial-mode-squeezed field could provide resilience to the type of mode-mismatch considered here. For practical implementation it would require a more detailed study and experimental demonstration. \\

The outline of this paper goes as follows. In Sec.~\ref{sec:ssm}, we go into the details of the model used to study the impact of spatial mode-mismatches, and we thoroughly analyze the results presented in Figs.~\ref{fig:ssm_ifo} and \ref{fig:ssm_fc} by using analytical expressions. In Sec.~\ref{sec:msm} we elaborate on the model used to study if the injection of squeezed states in multiple spatial modes could provide robustness to mode-mismatches, and the results presented above in Fig.~\ref{fig:msm_ifo} are further analysed. \\

\section{The effect of spatial mode-mismatches }
\label{sec:ssm}

We now go into the details behind the modeling of how mode-mismatches affects the quantum-noise-limited sensitivity of a squeezed-light-enhanced interferometric gravitational wave detector. 
Specifically, we start with the description of the optical setup in subsection~\ref{sec:setup}, and 
then in subsection~\ref{sec:framework}, we describe the general framework used to analyze the results. 
\textsc{Finesse}~\cite{Freise04, finesse20, freise10}---the numerical software that was used to produce the results---uses an equivalent method~\cite{phd.brown2015, Freise06}. A similar framework can also be found in Ref.~\cite{Zhang17}.
In the later subsections~\ref{sec:mmm_ifo},~\ref{sec:mmm_fc}, and~\ref{sec:mmm_sqz}, we look into mode-mismatches between the three components---the squeezer, the filter cavity, and the interferometer. \\

\subsection{The optical setup}
\label{sec:setup}

The optical setup used here is visualized in Fig.~\ref{fig:setup}, and is a simplified and idealized model of an Advanced LIGO detector~\cite{AdvancedLIGO15} with frequency dependent squeezed light injected through the dark port. The key parameters of the interferometer 
are listed in Table~\ref{tab:ifo}. 
The frequency dependent squeezing is realized by reflecting the squeezed field off a detuned over-coupled Fabry-Perot cavity. 
This cavity is frequently referred to as a filter cavity~\cite{Kimble02, Harms03, Chelkowski05}. 
The filter cavity considered in this work is a linear overcoupled 16 m long confocal optical cavity, based on the one proposed in~\cite{Evans2013} for near-term upgrade of Advanced LIGO. 
In this work, the input mirror is lossless, the end mirror is perfectly reflective, and we have assumed that the mirrors are much larger than the beam sizes so that clipping losses are negligible. 
The values used for cavity detuning and input mirror transmission were obtained by maximizing the broadband sensitivity between 10 Hz and 3 kHz. 
The radius of curvature for the two mirrors is chosen to make the higher-order modes resonant within the frequency band of interest, for the reason mentioned in the introduction. All the used filter cavity parameters are shown in Table~\ref{tab:fc}. \\

\begin{table}[h]
\begin{center}
\caption{The table shows the interferometer parameters that were used.}
\begin{tabular}{l c c }
  \hline
  \hline
  Symbol & Parameter & Value \\
  \hline
  $\lambda_0$ & Carrier wavelength & $1064$ nm\\
  $P_\mathrm{arm}$ & Arm cavity power & 0.74 MW  \\
  $P_\mathrm{bs}$ & Power on the beam splitter & 5.3 kW \\ 
  $L_\mathrm{arm}$ & Arm cavity lengths & 3994.5 m \\
  $m$ & Mass of test-mass mirrors & 40 kg \\
  $L_\mathrm{src}$ & Signal recycling cavity length & 57 m \\
  $T_\mathrm{srm}$ & SRM power transmission & 0.35 \\
   \hline
   \hline
\end{tabular}
\label{tab:ifo}
\end{center}
\end{table}

\begin{table}[h]
\begin{center}
\caption{The table shows the design parameters for the filter cavity used in Sec.~\ref{sec:ssm}.}
\begin{tabular}{l c c }
  \hline
  \hline
  Symbol & Parameter & Value \\
  \hline
  $L_\mathrm{fc}$ & Length & 16.0 m \\
  $R_\mathrm{C}$ & Mirror radius of curvature & 15.999 m \\
  $T_\mathrm{in}$ & Input mirror transmission & 61 ppm \\
  $R_\mathrm{in}$ & Input mirror reflection &  1-$T_{in}$ \\
  $R_\mathrm{end}$ & End mirror reflection &  1 \\
  FSR & Free spectral range & 9.37 MHz \\
  $\Delta /2\pi $ & Detuning & $ 46.18 $ Hz \\
  $\gamma_\mathrm{fc}^{}/2\pi$ & Half-width &  $ 45.49$ Hz \\
  $\delta_f$ & Mode-separation & $(1 + 4\times10^{-5}) \frac{\mathrm{FSR}}{2}$ Hz\\[0.15em]
  \hline
  \hline
\end{tabular}
\label{tab:fc}
\end{center}
\end{table}

We have three components to mode-mismatch to each other: the interferometer, the filter cavity, and the squeezer. The mode-mismatch between the interferometer and the filter cavity is generated by displacing a mode matching lens along the optical axis. For the squeezer component, \textsc{Finesse} allows us to freely specify the complex beam parameter of the field that is emitted, and we used this feature to control the mode matching of the squeezer. 

\subsection{The mathematical framework}
\label{sec:framework}

The spatial distribution of the field within the interferometer can be expanded in one common interferometer eigenbasis $U_n^\mathrm{IFO}(x, y, z)$. Specifically, the sideband field at $ \omega_0 \pm \Omega$ ($\omega_0$ is the carrier frequency of the laser) reads:
\begin{align}
	\hat{E}(\omega_0 \pm \Omega, x, y, z) = \sum_{n=0}^N c_{n} \hat{a}_{\omega_0\pm\Omega,n}U_n^\mathrm{IFO}(x, y, z)
	\end{align}
Here $\hat a_{\omega_0\pm\Omega,n}$ are the annihilation operators for the upper and lower sidebands of the $n$th mode, $c_n$ is the relative weight of the $n$th mode satisfying $\sum_{n=0}^\infty \abs{c_n}^2 = 1 $, $N$ denotes the number of modes included in the model,  $z$ is the coordinate along the optical axis, and $x$ and $y$ are the transverse coordinates. Similarly, the eigenbases of the filter cavity and the squeezer are denoted by $U_n^\mathrm{FC}$ and $U_n^\mathrm{SQZ}$, respectively. 
These are the three eigenbases used to describe the spatial distribution of the field within the optical setup. 
Which eigenbasis is used where is indicated by the background colors in Fig.~\ref{fig:setup}, and the red dots indicate where the basis changes take place. 
Scattering between modes labeled by different numbers $n$ occurs when changing basis from $U_n^\mathrm{SQZ}$ to $U_n^\mathrm{FC}$ and when changing basis from $U_n^\mathrm{FC}$ to $U_n^\mathrm{IFO}$, if the complex beam parameters of the bases are different. \\

In this paper, we use the two-photon formalism~\cite{Caves85, Schumaker85, CCM05} to model the quantum noise. 
In this formalism, the key quantitates are (i) the amplitude and phase quadrature operators which are defined as 
\begin{align}
	\hat{a}_1(\Omega) &= \frac{\hat{a}_{\omega_0 + \Omega} + \hat{a}_{\omega_0 - \Omega} ^\dagger}{\sqrt{2}} , \ 
	\hat{a}_2(\Omega) = \frac{\hat{a}_{\omega_0 + \Omega} - \hat{a}_{\omega_0 - \Omega}^\dagger}{\sqrt{2}\, i }
\label{eq:a_hat}
\end{align}
and (ii) the transfer matrix relating the quadrature operators of the fields at different locations. In our case, we care about higher-order modes where the quadrature operators can be represented in terms of a column vector of length $2 N$:
\begin{align}
	\mathbf{a}  = \bigoplus_{n=0}^N \mathbf{a}_n(\Omega)
\label{eq:a_bold}
\end{align}
with each pair of quadrature operators for mode $n$ being defined as
\begin{align}
	\mathbf{a}_n (\Omega) = 
	\begin{bmatrix} 
		\hat{a}_{1,n}(\Omega) & \hat{a}_{2,n}(\Omega) 
	\end{bmatrix}^\mathrm{T} .
\label{eq:an_bold}
\end{align}

The field that enters the interferometer can be related to the field entering the squeezer through 
\begin{align}
	\mathbf{a}_\mathrm{IFO}^{} = \mathcal{K}_2 \mathcal{T} \mathcal{K}_1 \mathcal{S}\,\mathbf{a}_\mathrm{SQZ}^{} ,
\label{eq:a_ifo}
\end{align}
Here, $\mathcal{S}$ is the squeezing matrix, $\mathcal{T}$ is the filter cavity transfer matrix, $\mathcal{K}_1$ describes the basis change from $U_n^\mathrm{SQZ}$ to $U_n^\mathrm{FC}$, and $\mathcal{K}_2$ describes the basis change from $U_n^\mathrm{FC}$ to $U_n^\mathrm{IFO}$. These matrices are described as follows. \\
 
The joint squeezing matrix $\mathcal{S}$ is given by the direct sum of the individual squeezing matrices for every spatial mode in the field:
\begin{align}
\mathcal{S} = \bigoplus_{n=0}^N \mathcal{S}_n.
\end{align}
The squeezing matrix $ \mathcal{S}_n$ for spatial mode $n$ is given by
\begin{equation}
	\begin{bmatrix*}[c]
		\cosh r_n + \sinh r_n\cos 2\varphi_n & \sinh r_n \sin 2\varphi_n \\[0.3em]
		\sinh r_n \sin 2\varphi_n & \cosh r_n- \sinh r_n \cos 2\varphi_n 
	\end{bmatrix*}, 
\label{eq:S}
\end{equation}
where $r_n$ and $\varphi_n$ are the squeeze factor and angle, respectively. 
In later subsections, the states in the fundamental mode are squeezed by 10 dB while all higher order modes contain pure vacuum states. That is, $r_0 = (2\log_{10}\mathrm{e})^{-1}$ and $r_n = 0$ for all $n>0$. The angle $\varphi_0$ is optimized such that the high-frequency shot noise is maximally reduced. The filter cavity then takes care of correctly rotating the squeezed states for the rest of the frequency components. \\

The matrix $\mathcal{K}$ describing a basis change between two spatial mode bases is given by
\begin{align}
{\cal K}= 
	\begin{bmatrix*}[c]
		 {\cal K}_{0,0} & \cdots & {\cal K}_{0,k} & \cdots & {\cal K}_{0,N} \\[0.3em]
		 \vdots & \ddots & \vdots & \ddots & \vdots \\[0.3em]
		 {\cal K}_{n,0} & \cdots & {\cal K}_{n,k} & \cdots & {\cal K}_{n,N} \\[0.3em]
		 \vdots & \ddots & \vdots & \ddots  & \vdots \\[0.3em]
		 {\cal K}_{N,0} & \cdots & {\cal K}_{N,k} & \cdots & {\cal K}_{N,N}
	\end{bmatrix*},
\label{eq:K}
\end{align}
where each entry ${\cal K}_{n,k}$ is a $2\times2$ matrix given by
\begin{align}
{\cal K}_{n,k} \equiv 
	\kappa_{nk}\begin{bmatrix*}[c]
		\cos \beta_{nk} & -\sin \beta_{nk} \\[0.3em]
		\sin \beta_{nk} & \cos \beta_{nk}
	\end{bmatrix*}.
\label{eq:Knk}
\end{align}
Here, $\kappa_{nk}$ is the coupling magnitude from mode number $k$ in the old basis to mode number $n$ in the new basis, and $\beta_{nk}$ is the corresponding coupling phase. \\

Expressed in the spatial basis $U_n^\mathrm{FC}$, the reflection off the filter cavity is given by 
\begin{align}
\mathcal{T}  = \bigoplus_{n=0}^N \mathcal{T}_n(\Omega),
\label{eq:T}
\end{align}
where the spatial mode $n$ undergoes a phase change specified by
\begin{align}
\mathcal{T}_n(\Omega) = \mathcal{A}_2 
	\begin{bmatrix*}[c]
		r_n^{}(\Omega) & 0 \\[0.3em]
		0 & r_n^{*}(-\Omega) 
	\end{bmatrix*} \mathcal{A}_2^{-1}.
\label{eq:Tn}
\end{align}
The transfer function for a sideband in spatial mode $n$ is given by
\begin{align}
	r_{n}^{}(\Omega) =   \frac{  \mathrm{e}^{-i\phi_n(\Omega)}-\sqrt{R_{\rm in}} } {\sqrt{R_{\rm in}} 
					\mathrm{e}^{-i\phi_n(\Omega)}  -1 },
\label{eq:rn}
\end{align}
where 
\begin{align}
	\phi_n(\Omega) = \left[ \frac{2 L}{c} \left(\Omega+\Delta \right) -  q_n\psi_\mathrm{rt}^{} \right]
\label{eq:phi_n}
\end{align}
and $R_\mathrm{in}$ is the input mirror power reflectivity, $\Delta$ is the cavity detuning, $L$ is the macroscopic cavity length, $c$ is the speed of light, $\psi_\mathrm{rt}^{}$ is the round-trip Gouy phase and $q_n$ is the order of the mode $n$. The matrix 
\begin{align}
\mathcal{A}_2 = \frac{1}{\sqrt{2}}
	\begin{bmatrix*}[r] 
		1 & 1  \\[0.3em]
		-i & i
	\end{bmatrix*} 
\end{align}
is used to transform the transfer function for the sidebands to that for the quadratures. \\

\subsection{Mode-mismatched interferometer}
\label{sec:mmm_ifo}
\begin{figure}[htb]
\begin{center}
\includegraphics[width=0.48\textwidth]{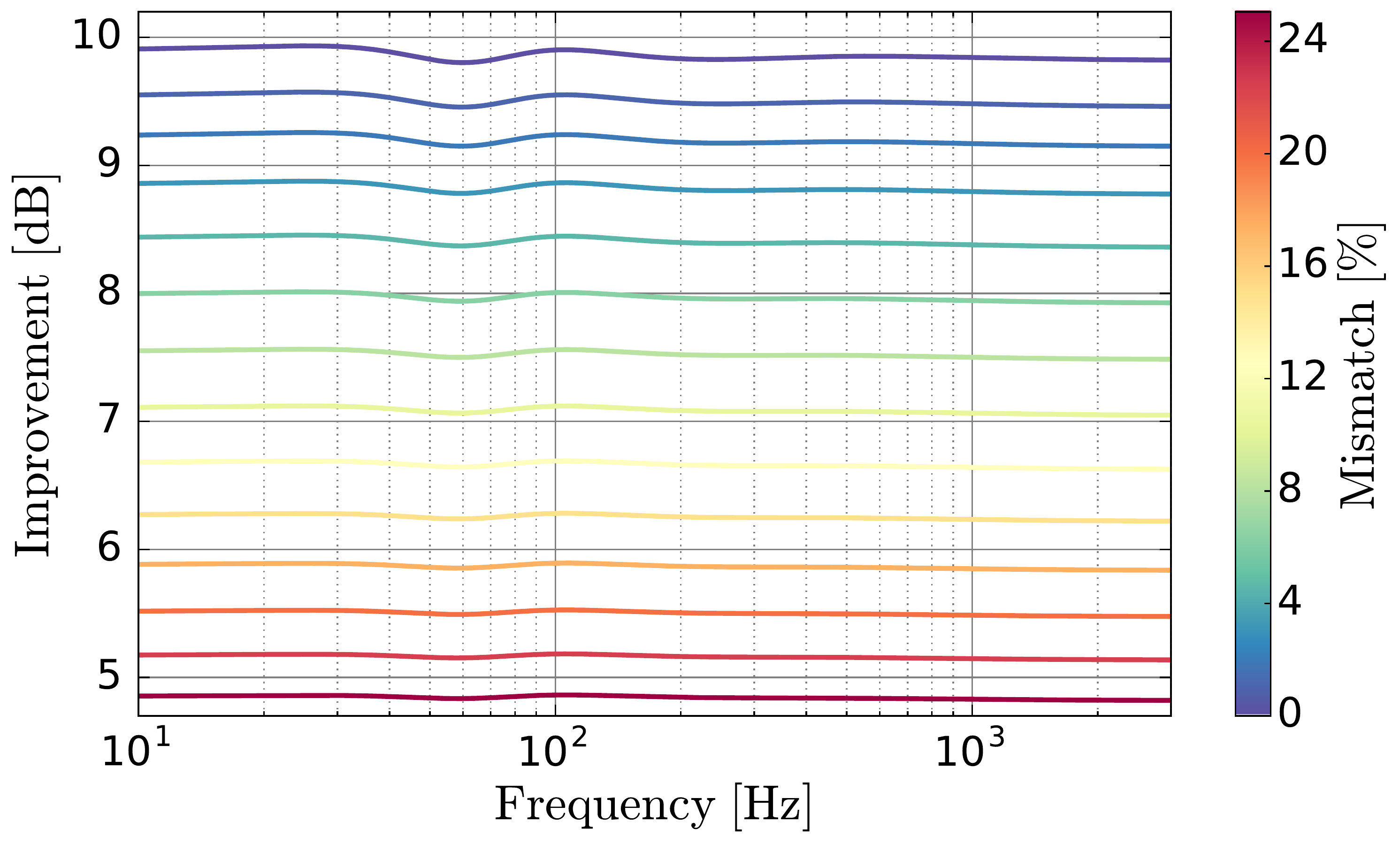}
\caption{The figure shows the effect of mode-mismatching the interferometer to the squeezer and the filter cavity. We see a broad-frequency band decrease in improvement that is similar to an optical loss.}
\label{fig:ssm_ifo_db}
\end{center}
\end{figure}

In this scenario, the interferometer is mode-mismatched to both the squeezer and the filter cavity, while the squeezer and the filter cavity are kept mode matched to each other. 
To generate this mode-mismatch, one of the lenses used to mode match the filter cavity to the interferometer is displaced along the optical axis. 
The resulting quantum-noise-limited sensitivity is shown in Fig.~\ref{fig:ssm_ifo}, while Fig.~\ref{fig:ssm_ifo_db} shows the same data but expressed in terms of improvement over the nonsqueezed case. 
The dip in improvement around 70 Hz demonstrates that one cannot achieve a perfect broad-frequency band noise reduction by using only one filter cavity~\cite{Kimble02}.
However, when operating with a tuned signal recycling cavity, as done here, one filter cavity still performs very well~\cite{Khalili07, Khalili10}. 
Since we are using realistic mirror losses inside the interferometer, the sensitivity improvement does not reach exactly 10 dB even when all three components are perfectly mode matched. 

\begin{figure}[htb]
\begin{center}
\includegraphics[scale=0.35]{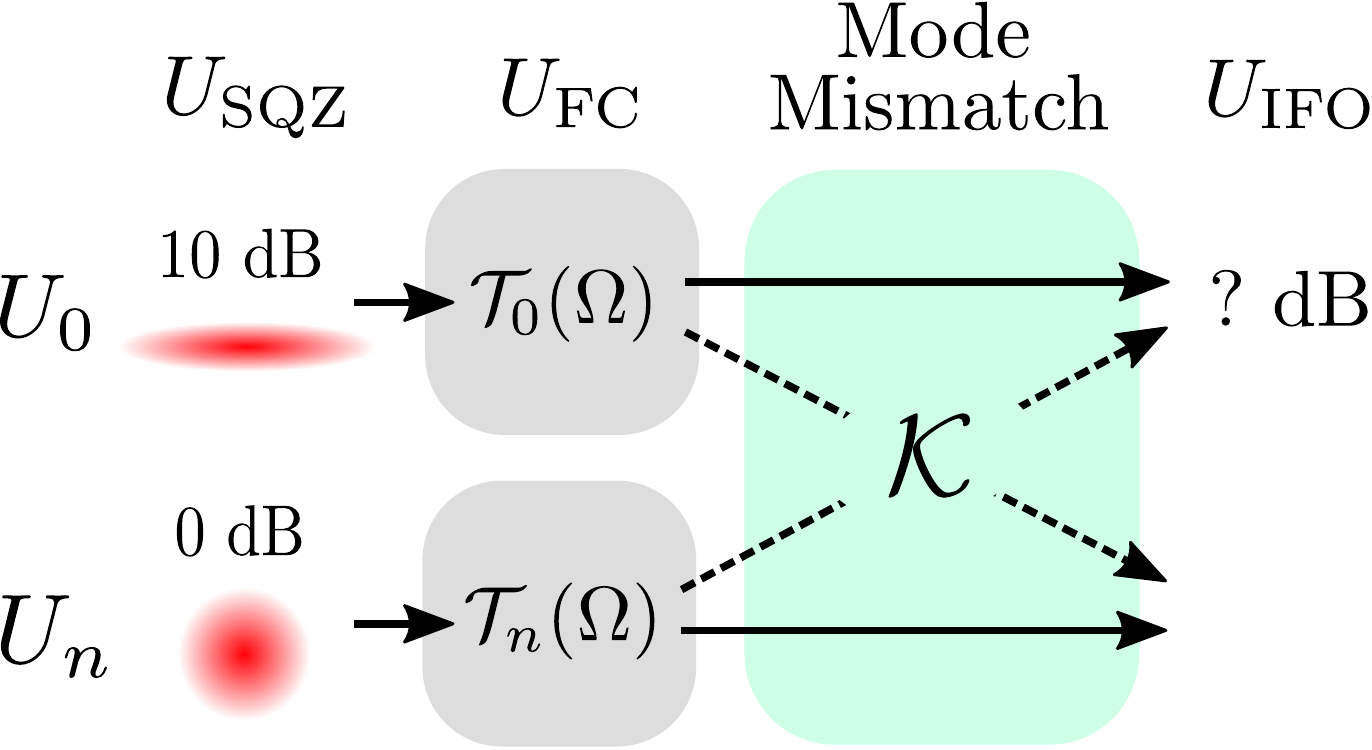}
\caption{The figure shows the effect of mode-mismatching the interferometer to the filter cavity and the squeezer. Initially, the vacuum noise in the spatial mode $U_0$ is squeezed by 10 dB, while the arbitrary higher-order mode $U_n$ contains pure vacuum noise. The noise fields in the two spatial modes mix after being subjected to the frequency and mode dependent rotation of the squeeze angle when reflected off the filter cavity.}
\label{fig:explained_IFO}
\end{center}
\end{figure}

The reason for the broad-frequency band squeezing degradation is best explained by using the analytics developed above. 
Since the squeezer and the filter cavity are mode matched, and assuming that the self-coupling phases in equation~\ref{eq:Knk} are $\beta_{kk} = 0$, the basis change matrix $\mathcal{K}_1$ in equation~\ref{eq:a_ifo} becomes the identity matrix. 
This assumption does not reduce the generality as any self-coupling phase could be compensated for by adjusting the initial squeeze angle. 
Equation~\ref{eq:a_ifo}, describing the quantum field injected into the interferometer, is then reduced to 
\begin{align}
\mathbf{a}_\mathrm{IFO}^{} = \mathcal{K} \mathcal{T} \mathcal{S}\, \mathbf{a}_\mathrm{SQZ}^{},
\label{eq:Ma_ifo}
\end{align}
which is visualized in Fig.~\ref{fig:explained_IFO}. 
The only frequency dependent process that the field undergoes is the interaction with the filter cavity, which is described by equation~\ref{eq:T}. When this process takes place, all the squeezed states are in the fundamental mode and therefore undergo the correct rotation $T_0(\Omega)$. 
The phase changes of the pure vacuum states in the higher-order modes $T_n(\Omega)$ are unimportant, as these just rotate circular symmetric probability distributions around their symmetry axes.

The mode-mismatch-induced basis change $\mathcal{K}$ makes the fundamental mode exchange some squeezed states for pure vacuum states with the higher-order modes. This makes the fundamental mode of the interferometer eigenbasis less squeezed for all frequencies, and has the same effect as an optical loss. That is, for small coupling coefficients $\kappa_0$, where
\begin{align}
\kappa_0^2 = \sum_{n=1}^N \kappa_{0n}^2
\end{align}
is the total power coupling magnitude for scattering away from the fundamental mode, the quantum noise in the interferometer scales as
\begin{align}
 ({1-\kappa_0^2})\mathrm{e}^{-2r_0} + \kappa_0^2.
\end{align}

\subsection{Mode-mismatched filter cavity}
\label{sec:mmm_fc}

Just as above, the filter cavity is spatially mode-mismatched to the interferometer, but here the squeezer is kept mode matched to the interferometer instead of to the filter cavity. 

\begin{figure}[htb]
\begin{center}
\includegraphics[scale=0.35]{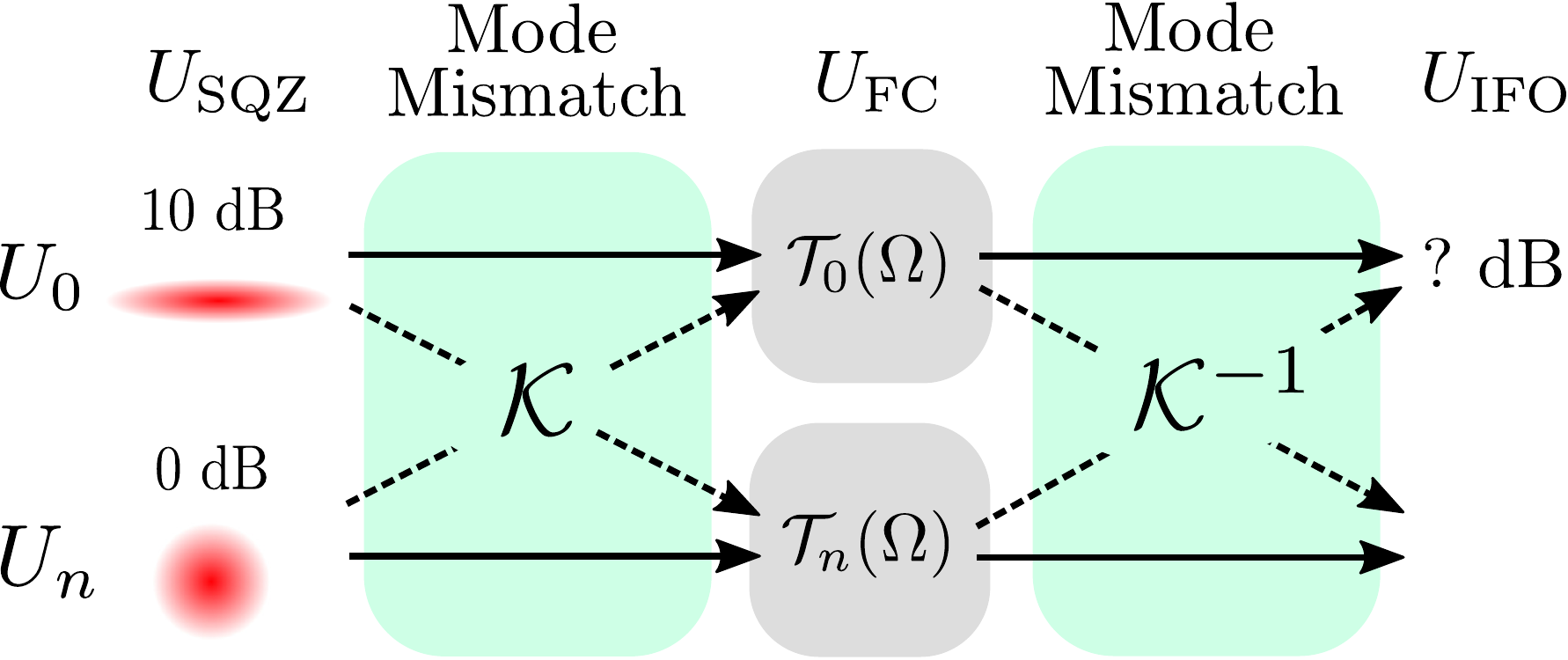}
\caption{The figure shows the effect of mode-mismatching the filter cavity to the squeezer and the interferometer. The noise fields in the two spatial modes mix twice, with a frequency and mode dependent rotation of the squeeze angle in between due to the filter cavity. Since the squeezer and the interferometer are mode matched, the two mixing operations are the inverse of each other.}
\label{fig:explained_FC}
\end{center}
\end{figure}

In this case, there are nontrivial spatial basis changes before and after the filter cavity that give rise to couplings between different spatial modes.
Since the squeezer and the interferometer are mode matched to each other, the second basis change is the inverse of the first, thus, equation~\ref{eq:a_ifo} becomes
\begin{align}
	\mathbf{a}_\mathrm{IFO}^{} = \mathcal{K}^{-1} \mathcal{T} \mathcal{K} \mathcal{S}\, \mathbf{a}_\mathrm{SQZ}^{}. 
\label{eq:mmm_fc_a_ifo}
\end{align}
This process is visualized in Fig.~\ref{fig:explained_FC}. \\

\begin{figure}[t]
\begin{center}
\includegraphics[width=0.48\textwidth]{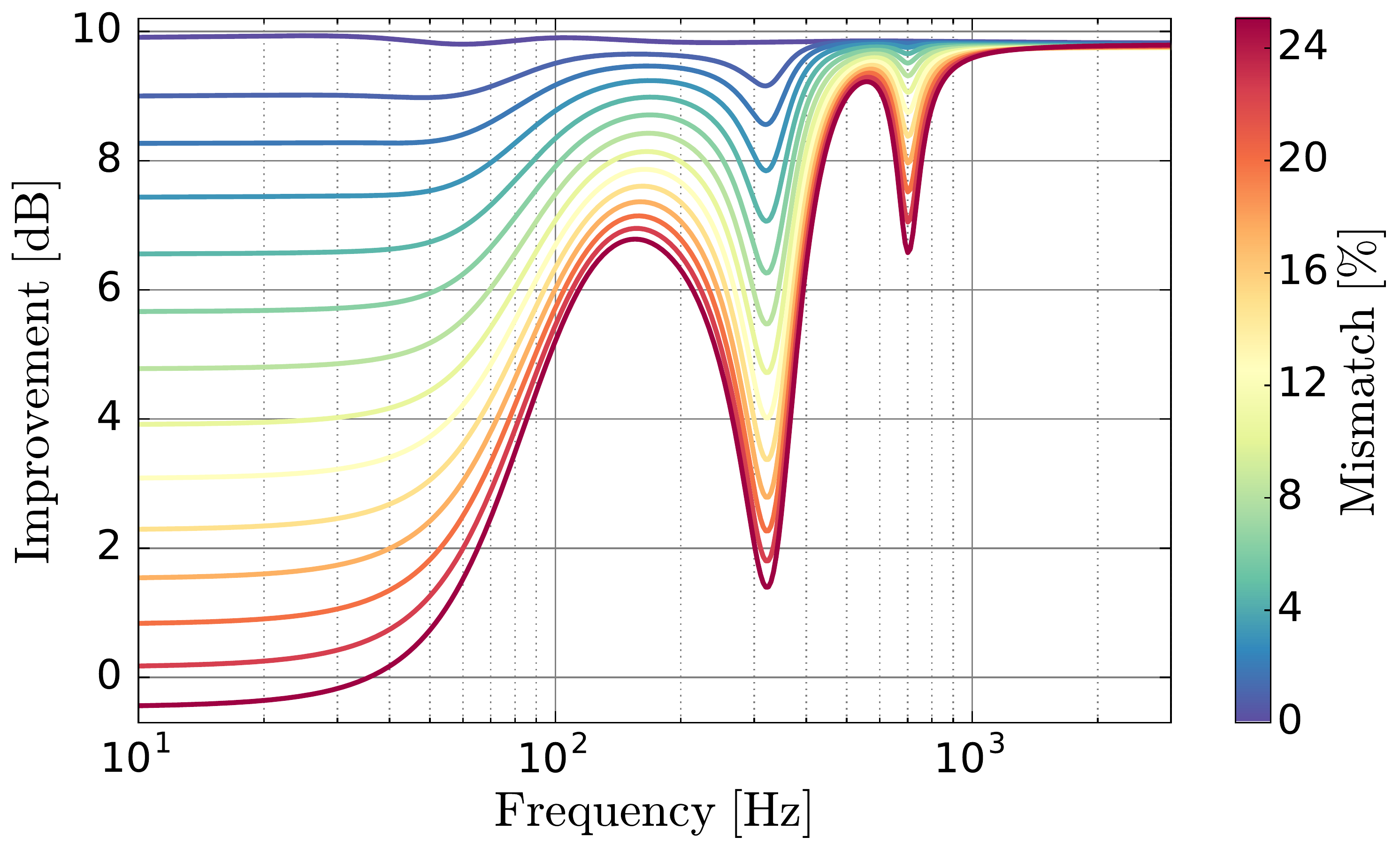}
\caption{The figure shows the effect of mode-mismatching the filter cavity to the squeezer and the interferometer. Due to coherent mode-scattering, we see squeezing degradations around the resonance frequencies for the involved spatial modes.}
\label{fig:ssm_fc_db}
\end{center}
\end{figure}

Due to the mode-mismatch $\mathcal{K}$ between the squeezer and the filter cavity, the field incident on the filter cavity input mirror has a part of its squeezed states located in higher-order modes.
If these higher-order modes experience phase shifts different from the phase shift of the fundamental mode when reflected off the filter cavity (i.e., if $T_n(\Omega) \neq T_0(\Omega)$), then the mode-mismatch between the filter cavity and the interferometer, $\mathcal{K}^{-1}$, enables for these now wrongly rotated squeezed states to mix back in with the squeezed states in the fundamental mode. 
If the wrongly rotated states are antisqueezed, this coherent scattering process is worse than an optical loss. 
In Figs.~\ref{fig:ssm_fc} and~\ref{fig:ssm_fc_db}, this coherent scattering effect can be seen in two different regions:
at low frequencies, where the fundamental mode is nearly resonant while the higher-order modes are off resonance, and at about 300 Hz and 700 Hz where the second-order and fourth-order modes are resonant while the fundamental mode is not. 
The reason that the second-order and fourth-order modes show up is that the mode-mismatch was generated by offsetting the waist size and displacing the waist position of the beam, which only generates nonzero couplings between modes with even mode-order spacing. 
Since the couplings decrease with increasing mode-order spacing, we only included modes up to order four in our simulations. 

For a small mode-mismatch, and for the worst case higher-order-mode rotations, the quantum noise in the interferometer scales as
\begin{align}
\mathrm{e}^{-2r}  + 4(1-\mathrm{e}^{-2r})\kappa_0^2.
\end{align}
See Appendix~\ref{sec:scaling} for a derivation of this formula. For large squeeze magnitudes, this is a factor of 2 worse than the effect of a corresponding optical loss.
It should be mentioned that the filter cavity was deliberately designed to have this small mode spacing so that we could see the effect of higher-order mode resonances. If this 16 m filter cavity would be implemented in LIGO, it would be designed such that the higher-order modes are resonant well outside the frequency range of interest. However, this might not be possible for much longer filter cavities, e.g., as proposed for the Einstein Telescope~\cite{Punturo10a}.

For high frequencies, neither the fundamental mode nor the higher-order modes are resonant, thus $T_n(\Omega)=T_0(\Omega)$, and the squeezed field is consequently unaffected by this mode-mismatch.

\subsection{Mode-mismatched squeezer}
\label{sec:mmm_sqz}

\begin{figure}[tbh]
\begin{center}
\includegraphics[scale=0.35]{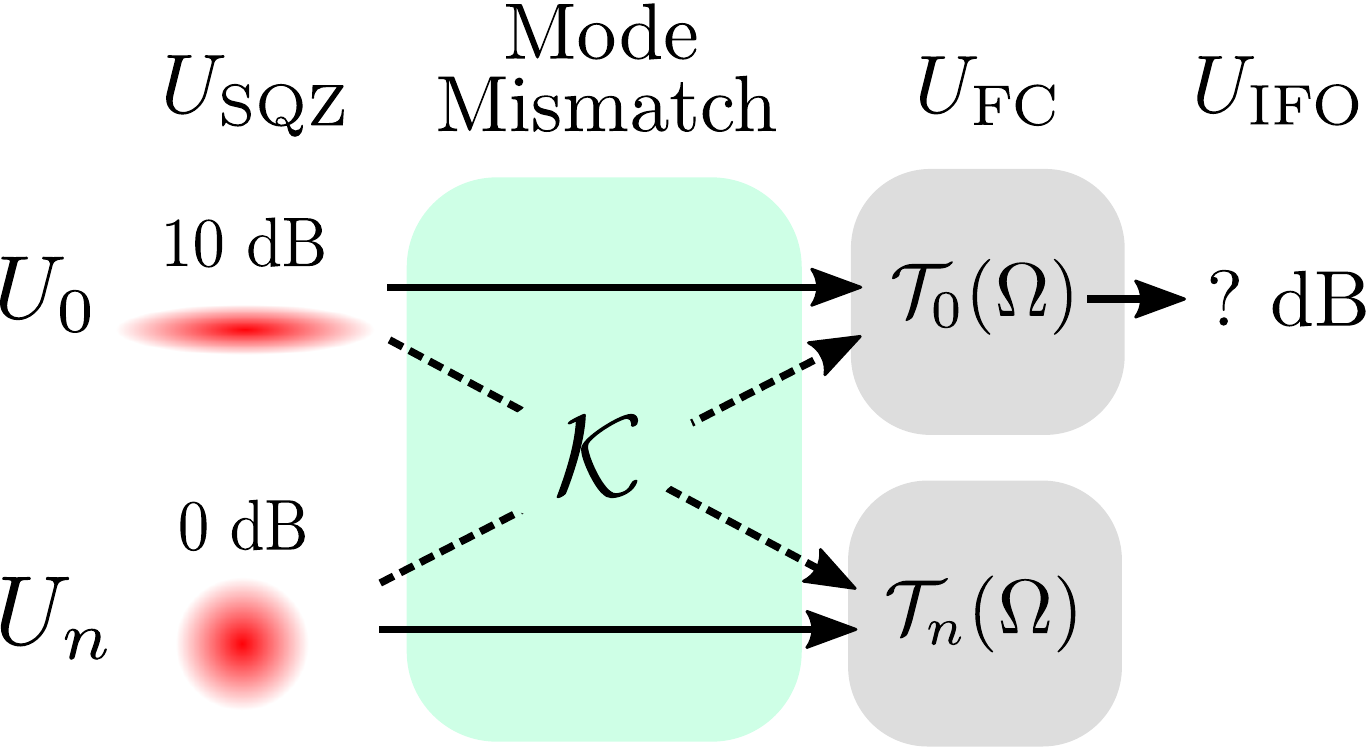}
\caption{The figure shows the effect of mode-mismatching the Squeezer to the FC and the Interferometer. The noise fields in the two spatial modes mix before being subjected to the frequency and mode dependent rotation of the squeeze angle when reflected off the FC.}
\label{fig:explained_SQZ}
\end{center}
\end{figure}

Here we consider the case where the squeezer is mode-mismatched to both the filter cavity and the interferometer, while the last two are kept mode matched to each other. This means that the basis change between the squeezer and the filter cavity generally has nonzero couplings between different spatial modes, while the matrix performing the basis change in between the filter cavity and the interferometer becomes the identity matrix. Thus, equation~\ref{eq:a_ifo} becomes
\begin{align}
\mathbf{a}_\mathrm{IFO}^{} = \mathcal{T} \mathcal{K} \mathcal{S} \, \mathbf{a}_\mathrm{SQZ}^{},
\end{align}
which is visualized in Fig.~\ref{fig:explained_SQZ}. The effect is the same in Sec.~\ref{sec:mmm_ifo}, thus the result can be seen in Figs.~\ref{fig:ssm_ifo} and~\ref{fig:ssm_ifo_db}. 
In contrast to the case in Sec.~\ref{sec:mmm_ifo}, there are indeed squeezed states in the higher-order modes that have incorrect rotations due to the filter cavity. But since these are not allowed to couple back to the fundamental mode again, this does not contribute to any extra quantum noise. 

\section{Robustness to mode-mismatches through squeezed higher-order modes}
\label{sec:msm}

In this section, we show that the injection of squeezed states in multiple spatial modes potentially can provide robustness to mode-mismatches. 
This requires that the initial orientation of the squeezing ellipses can be independently optimized for each spatial mode, which would be challenging to achieve in practice due to the degenerate resonance conditions of the second order modes. 
Further, the field from three different squeezers would have to be superimposed into one by using mode-selecting cavities. 

In subsection~\ref{sec:msm_ifo} the mode-mismatched interferometer is revisited (see Sec.~\ref{sec:mmm_ifo}), but this time three spatial modes are squeezed instead of just the fundamental mode. Subsection~\ref{sec:msm_principle} provides a simple analytic test of the principle of using multiple squeezed modes---it was not rejected.

\subsection{Mode-mismatched interferometer}
\label{sec:msm_ifo}

The same mode-mismatch is considered as in Sec.~\ref{sec:mmm_ifo}, that is, the interferometer is mode-mismatched to the filter cavity and the squeezer, while the filter cavity and the squeezer are kept mode matched to each other. Therefore, equation~\ref{eq:Ma_ifo} applies here as well, but with some alterations to the squeezing matrix $\mathcal{S}$ and to the filter cavity transfer matrix $\mathcal{T}$, as described below. 

We squeezed the Hermite-Gaussian modes $\mathrm{HG}_{02}$ and $\mathrm{HG}_{20}$, in addition to the fundamental mode, as these two second order modes have the strongest couplings to the fundamental mode, as mentioned in Sec.~\ref{sec:mmm_fc}. All three states are squeezed by 10 dB. 
The two extra modes are labeled $n=1$ and $n=2$, thus, the squeeze magnitudes in the squeezing matrix $\mathcal{S}$ (equation~\ref{eq:S}) becomes $r_n  = (2\log_{10}\mathrm{e})^{-1}$ for $n\in \{0,1,2 \}$, and $r_n  = 0$ for $n>2$. 
Further, for each level of mode-mismatch the initial squeeze angles $\varphi_n$ for $n\in\{0,1,2\}$ are independently optimized to maximize the sensitivity (or equivalently, to minimize the quantum noise). 
This optimization is needed to correctly compensate for the phases $\beta_{0k}$, $k\in\{1,2 \}$, that are picked up when the squeezed higher order modes couple into fundamental mode due to the mode-mismatch-induced basis change $\mathcal{K}$ (equation~\ref{eq:K}). 

To acquire the optimal frequency dependent rotation for the squeezed states in all three spatial modes, the filter cavity was made critical by changing the radius of curvature of the two filter cavity mirrors to 16 m. 
This gives a round-trip Gouy-phase of $\pi$, hence, the second order modes have the same resonance condition as the fundamental mode, and therefore pick up the same phase shift modulo $2\pi$ when subjected to filter cavity transfer matrix $\mathcal{T}$. 
This can be seen by setting $\psi_\mathrm{rt} = \pi$, $q(0) = 0$ and $q(1) = q(2) = 2$ in equation~\ref{eq:phi_n}. \\ 

The results for two different levels of mode-mismatches are shown in Fig.~\ref{fig:msm_ifo}, and are presented in terms of sensitivity improvement over the no-squeezing case. The figure also includes the corresponding traces from subsection~\ref{sec:mmm_ifo} for comparison. 
One can see that for 5 \% mode-mismatch the sensitivity is increased with about 1.5 dB compared to the case when only the fundamental mode is squeezed, and that most of the mode-mismatch-induced squeezing degradation is recovered by squeezing the two extra spatial modes. There are two reasons for this: \begin{enumerate}
	\item[(i)]  In the previous section, pure vacuum states from the second-order modes mixed in with the squeezed states in the fundamental mode due to the mode-mismatch. Now, correctly rotated squeezed states mix in instead.
	\item[(ii)] The couplings between the fundamental mode and the higher-order modes that carry pure vacuum states are small for this level of mode-mismatch. 
\end{enumerate}

For the larger mode-mismatch of 15 \%, the sensitivity gain is also larger---about 3 dB. This is because the coupling magnitudes between the fundamental mode and the second-order modes have increased. 
However, the sensitivity does not rise to around the mode matched case, as the fundamental mode has significant couplings to pure-vacuum-state-carrying higher-order modes. 
The results show that squeezing the two extra spatial modes provide robustness to this particular mode-mismatch in our model.

\subsection{Test of principle}
\label{sec:msm_principle}

In this subsection we provide a test of principle for multi-spatial-mode squeezing by injecting two squeezed quantum fields into a Mach-Zehnder interferometer. 
The test originated from the idea of testing if the benefits of squeezing higher-order modes could be downgraded or even rejected, if we allow propagations and scatterings that are more general in nature than the ones studied in the previous subsection. 

\begin{figure}[t]
\begin{center}
\begin{subfigure}[h]{0.235\textwidth}
       \centering
       \includegraphics[scale=0.1]{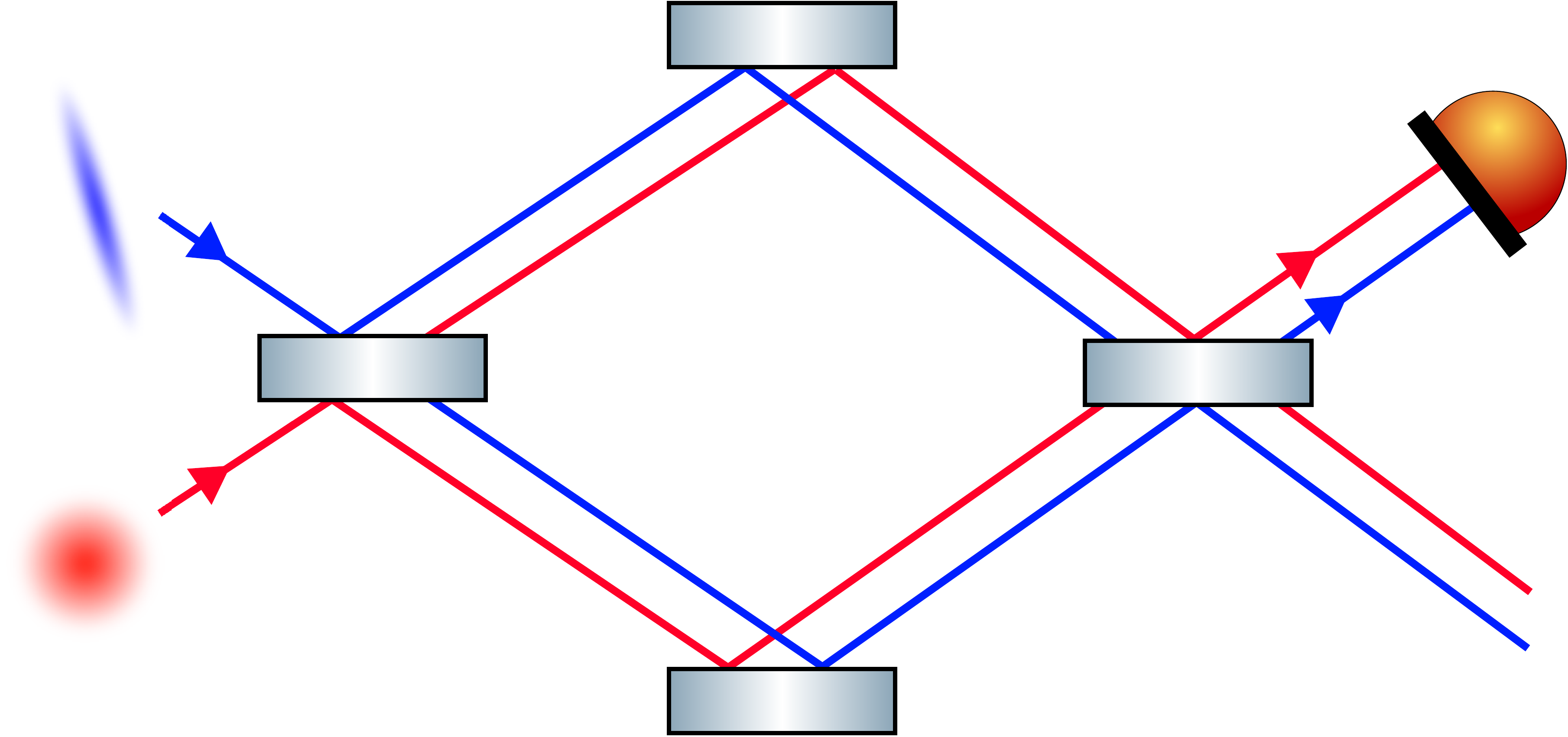}
       \label{fig:mz2}
\end{subfigure}
\begin{subfigure}[h]{0.235\textwidth}
      \centering
       \includegraphics[scale=0.1]{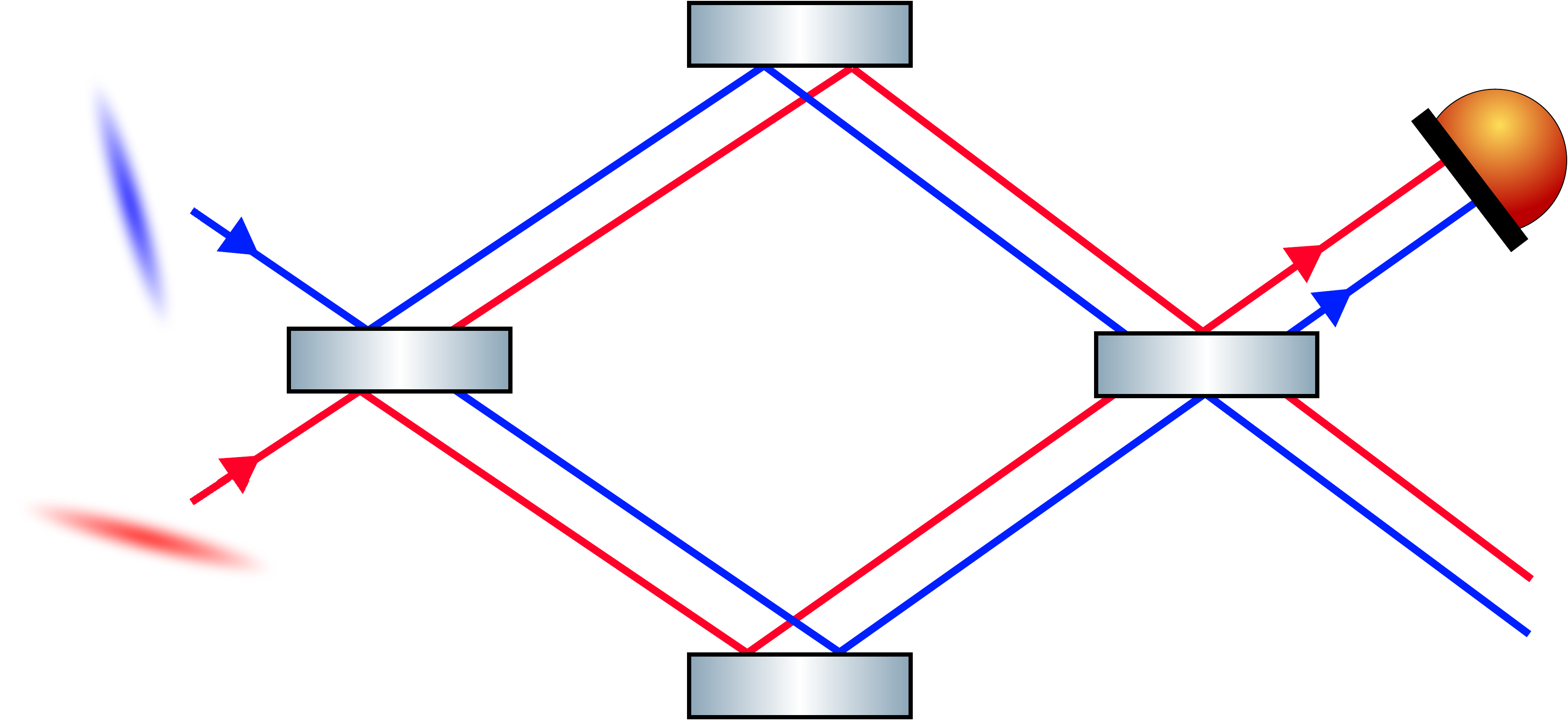}
       \label{fig:mz3}
\end{subfigure}
\caption{The figure shows Mach-Zehnder interferometers used to mix two quantum fields. In the left figure a squeezed vacuum field is mixed with pure vacuum, and in the right figure two independent squeezed vacuum fields are mixed. The photo detector indicates that the upper output path is the one of interest.}
\label{fig:mz}
\end{center}
\end{figure}

The optical setup is shown in Fig.~\ref{fig:mz} and consists of two squeezers---one for each incoming field---and two mixing points with a generic propagation in between. The test was performed as follows:

\begin{enumerate}
\item[(i)]
Various parameters of the system are independently assigned randomized values within realistic and physically valid intervals. These parameters are: the beam splitters' reflection coefficients and microscopical offsets along their surface normals; the macroscopical and microscopical propagation phases; and the readout quadrature. Here, microscopical refers to distances smaller than the carrier wavelength, and macroscopical refers to distances of any magnitude, but of integer multiples of the carrier wavelength. 
\item[(ii)] 
The upper input field is squeezed by 10 dB and the lower input field remains pure vacuum, as seen in the left part of Fig.~\ref{fig:mz}. The initial squeeze angle is optimized to yield maximum squeezing in the upper output path in the readout quadrature.
\item[(iii)]
The second squeezer is switched on so that both fields are squeezed by 10 dB, as seen in the right part of Fig.~\ref{fig:mz}. The initial squeeze angle for the lower field is then also optimized to yield maximum squeezing in the upper output path in the readout quadrature. 
\item[(iv)]
Repeat 10,000 times. 
\end{enumerate}

\begin{figure}[t]
\begin{center}
\includegraphics[width=0.48\textwidth]{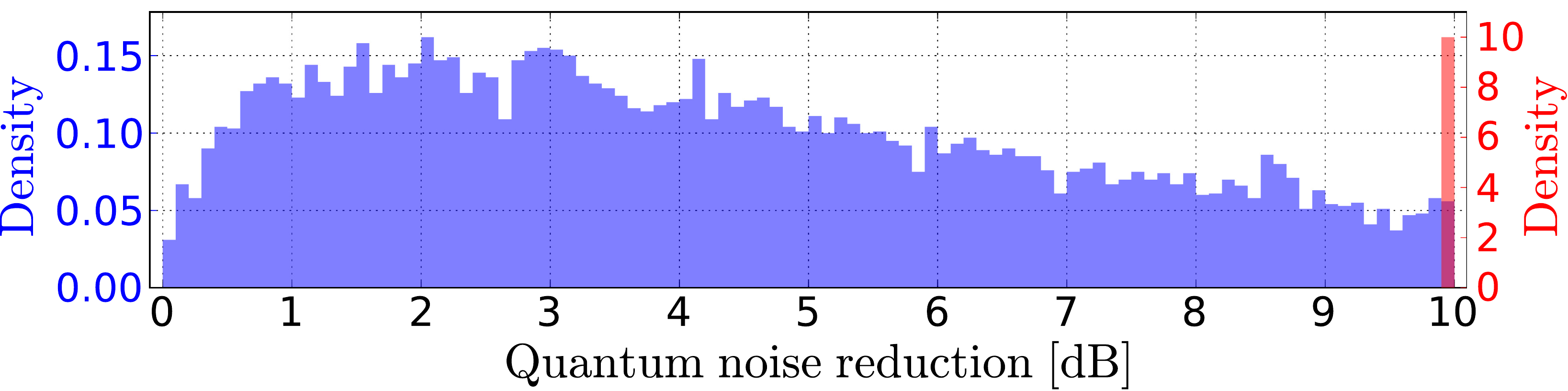}
\caption{The figure shows the "probability density" for obtaining a certain quantum noise reduction when mixing a squeezed vacuum with pure vacuum (blue) and when mixing two squeezed vacuum fields (red). The squeeze angles have, for both distributions, been optimized to minimize the noise.}
\label{fig:mz_hist}
\end{center}
\end{figure}

The result is shown in Fig.~\ref{fig:mz_hist}. The blue distribution is obtained with one squeezed field in step (ii), and the red bar is the result obtained in step (iii), when both fields are squeezed. Thus, for any set of random parameter values, we can always obtain 10 dB of squeezing as long as we can independently optimize the two initial squeeze angles. \\

The rest of this subsection is focused on describing the model that was used in more detail. 

The system can be described by the framework from Sec.~\ref{sec:framework}, with $N=1$ as there are only two fields in this setup. The upper (lower) field, and the operations acting on the upper (lower) field, are everywhere in the setup labeled by $n=0$ ($n=1$). The relation between the output fields and the input vacuum fields is given by equation~\ref{eq:a_ifo}, however, the transfer matrices $\mathcal{K}_1$, $\mathcal{K}_2$ and $\mathcal{T}$ are modified as follows.\\

Each lossless beam splitter can be represented by 
\begin{align}
	\mathcal{K}_i = 
	\begin{bmatrix}
		r_i \cos \beta_i & -r_i \sin \beta_i & t_i & 0 \\[0.3em]
		r_i \sin \beta_i & r_i \cos \beta_i  & 0 & t_i \\[0.3em]
		t_i & 0 & -r_i  \cos \beta_i & -r_i \sin \beta_i \\[0.3em]
		0 & t_i & r_i  \sin \beta_i & -r_i  \cos \beta_i
	\end{bmatrix}
\end{align}
where $r_i \in [0.7,1]$ is the reflection coefficient, $t_i$ is the transmission coefficient satisfying $t_i^2 = 1-r_i^2$, and $\beta_i \in [-\pi,\, \pi]$ is the phase shift due to the displacement of the beam splitter along its surface normal. \\

The propagation $\mathcal{T}$ consists of two independent paths of lengths $D_n = L_n + \delta L_n$, where $| \delta L_n | < \lambda_0$ and $L_n = k_n \lambda_0$ with $k_n \in \mathbb{N}$. Thus, the transfer matrices for paths $n=0,1$ are given by
\begin{align}
	\mathcal{T}_n(\Omega) = \mathrm{e}^{-i\theta_n}
	\begin{bmatrix}
	\cos\phi_n & \sin \phi_n \\[0.3em]
	-\sin \phi_n & \cos \phi_n
	\end{bmatrix}.
\end{align}
Here,
\begin{align}
	\theta_n = \frac{\Omega L_n}{c} \in [0, \pi]
\label{eq:macro}
\end{align}
is the phase picked up due to the macroscopical length $L_n$, and 
\begin{align}
	\quad \phi_n = \frac{\omega_0 \delta L_n}{c} \in [-\pi,\pi].
\label{eq:micro}
\end{align}
is the phase shift induced by the microscopical length $\delta L_n$.

\section{A more realistic Advanced LIGO model}
\label{sec:new_file}

To get a hint of how mode-mismatches inside the interferometer affect the multi-spatial-mode squeezed field, we here consider a \textsc{Finesse} model of an advanced LIGO detector that includes small mode-mismatches between the cavities inside the interferometer. 

There are two important differences compared to the model described in Sec.~\ref{sec:ssm}. The first one is that the asymmetries between the two transverse spatial directions are included in the model, which gives rise to mode-mismatches that are small, but not negligible.
These asymmetries show up because of nonzero angles of incidence in combination with spherical mirrors.
The second important difference is that an Advanced LIGO output mode cleaner has been added to the model. The reason for this is that some fraction of the coherent laser power is in higher-order modes due to the internal mode-mismatches. 
Without the output mode cleaner, higher-order modes of the quantum field are allowed to beat with the higher-order modes of the coherent carrier field. This creates noise that would not be present with the output mode cleaner included. \\

\begin{figure}[htb]
\begin{center}
\includegraphics[width=0.48\textwidth]{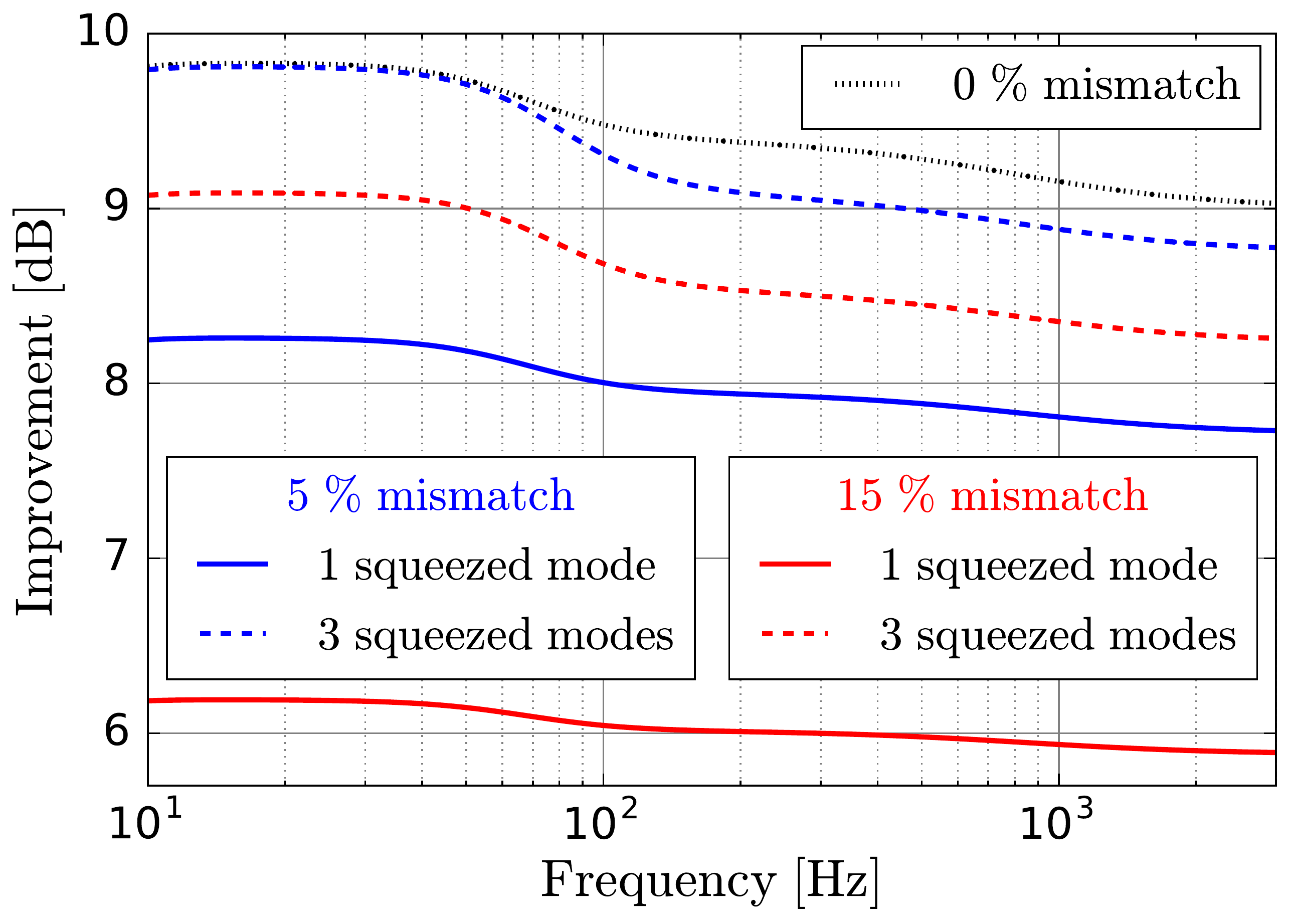}
\caption{The figure shows the improvement in quantum-noise-limited sensitivity over the nonsqueezed case, both for a single squeezed spatial mode (solid lines) and for multiple squeezed spatial modes (dashed lines). The blue and red traces are for mode-mismatch levels between the interferometer and the filter cavity of 5 \% and 15 \%, respectively. The squeezer is kept mode matched to the filter cavity. The black dotted trace indicates the improvement when the filter cavity and the output mode cleaner are near perfectly mode matched.}
\label{fig:new_file}
\end{center}
\end{figure}

The experiment was performed by mode-mismatching the filter cavity to the output mode cleaner by varying the position of a mode matching lens along the optical axis. This mode matching lens is located between the filter cavity and the injection point for the squeezed field. 
The squeezer was kept mode matched to the filter cavity. 
We computed the quantum-noise-limited-sensitivity in the frequency band of interest for two levels of mode-mismatches. This was done both for a squeezer that emits one and three squeezed spatial modes. 
The resulting improvements over the no-squeezing case are shown in Fig.~\ref{fig:new_file}. The behavior at low frequencies is identical to the result obtained with the simpler model considered in Sec.~\ref{sec:msm}. At high frequencies, the squeezed field experiences a slightly larger degradation, which mainly seems to be due to the output mode cleaner, however, further investigation is needed to conclude this.

Moreover, we can conclude that the internal mode-mismatches included in this model are too small to give rise to any large effects. Future work aims at systematically study the impact of internal mode-mismatches due to, e.g., thermal lensing. 

\section{Conclusions}

In this paper, we have quantified and described how squeezed-light-enhanced interferometric gravitational-wave detectors are affected by spatial mode-mismatches between the interferometer, the filter cavity, and the squeezer. 
We have shown that spatial mode-mismatches potentially can cause significantly larger squeezing degradations than a pure optical loss, if multiple mode-mismatches allow squeezed states to coherently scatter back and forth between the fundamental mode and higher order modes. 
We can conclude that even with relatively large mode-mismatches, the injection of frequency dependent squeezed light is beneficial in our model.

Further, we have shown that the injection of a field with squeezed states, not only in the fundamental mode, but also in the second-order Hermite-Gaussian modes $\mathrm{HG}_{02}$ and $\mathrm{HG}_{20}$, potentially can provide resilience to spatial mode-mismatches. This scheme requires independent optimization of the squeeze angles for all three involved spatial modes, which poses a big challenge for any potential real-world implementation.

Further studies of how combinations of external and intra-interferometer spatial mode-mismatches affect the performance of squeezed light are needed to better understand how squeezed light would perform in gravitational wave detectors. 

\section{Acknowledgements}
\label{sec:acknowledgements}

The authors would like to thank Matthew Evans and Rana Adhikari for useful discussions. This work was supported by the Science and Technology Facilities Council Consolidated Grant (number ST/N000633/1).
D.~T\"oyr\"a was supported by the People Programme (Marie Curie Actions) of the European Union's Seventh Framework Programme FP7/2007-2013/ (PEOPLE-2013-ITN) under REA grant agreement n° [606176]. 
H.~Miao was supported by UK STFC Ernest Rutherford Fellowship (Grant No. ST/M005844/11).
M.~Davis was funded by the United States National Science Foundation via grant number PHY-1460803 to the University of Florida Gravitational Physics IREU program.

\appendix

\section{Noise scaling of the coherent scattering effect}
\label{sec:scaling}

In this section we derive how the noise due to the coherent scattering effect scales with the coupling coefficient. 
We use a simplified version of the system considered in Sec.~\ref{sec:mmm_fc} where the filter cavity is mode-mismatched to the interferometer and the squeezer, while the squeezer and the interferometer are kept mode matched.
Here, we only use two fields, i.e., $N = 1$ in the mathematical framework in Sec.~\ref{sec:framework}. The relation between the output field and the input field is given by equation~\ref{eq:mmm_fc_a_ifo}, but where the matrices are simplified.

Only one of the two fields is squeezed, thus, the squeezing matrix can be written as
\begin{align}
	S = 
	\begin{bmatrix*}[l]
		\mathrm{e}^{r} & 0 & 0 & 0 \\[0.3em]
		0 & \mathrm{e}^{-r} & 0 & 0 \\[0.3em]
		0 & 0 & 1 & 0 \\[0.3em]
		0 & 0 & 0 & 1
	\end{bmatrix*}.
\end{align}
The scattering matrix is given by 
\begin{align}
	\mathcal{K} = 
	\begin{bmatrix}
	\cos \kappa & 0 & -\sin \kappa & 0 \\[0.3em]
	0 & \cos \kappa & 0 & -\sin \kappa \\[0.3em]
	\sin \kappa & 0 & \cos \kappa & 0 \\[0.3em]
	0 & \sin \kappa & 0 & \cos \kappa
	\end{bmatrix} ,
\end{align}
where $\sin \kappa$ is the coupling between the two fields. For the propagation, only the relative phase shift between the two fields is of importance, hence it can be represented by the matrix
\begin{align}
	\mathcal{T} = 
	\begin{bmatrix}
	1 & 0 & 0 & 0 \\[0.3em]
	0 & 1 & 0 & 0 \\[0.3em]
	0 & 0 & \cos \phi & -\sin \phi \\[0.3em]
	0 & 0 & \sin \phi & \cos \phi
	\end{bmatrix} ,
\end{align}
where $\phi$ is the relative phase shift. Assuming we are squeezing the readout quadrature, the noise is proportional to the element $\mathcal{M}(2,2)$, where
\begin{align}
\mathcal{M} &= \mathcal{K}^{-1} \mathcal{T} \mathcal{K} \mathcal{S} \big( \mathcal{K}^{-1} \mathcal{T} \mathcal{K} \mathcal{S} \big)^\mathrm{T} \\
	&= \mathcal{K}^{-1} \mathcal{T} \mathcal{K} \mathcal{S}^2 \mathcal{K}^{-1} \mathcal{T}^\mathrm{T} \mathcal{K}.  \\
\end{align}
Assuming the coupling magnitude $\sin\kappa$ is small, then 
\begin{align}
\mathcal{M}(2,2) = \mathrm{e}^{-2r} - 2 \kappa^2 \mathrm{e}^{-2r} \big(\mathrm{e}^{2r} - 1 \big) \big( \cos \phi  - 1 \big) + \mathcal{O}(\kappa^3) .
\end{align}
Thus, the worst case scenario is if the propagation gives rise to a relative phase shift between the two fields of $\phi = \pi$, in which case the noise arising due to the coherent scattering effect scales as
\begin{align}
\mathrm{e}^{-2r} + 4\kappa^2\big( 1-\mathrm{e}^{-2r} \big) +  \mathcal{O}(\kappa^3) .
\end{align}
For large squeeze magnitudes, this is a factor of two worse than if these two scattering points would have been exchanged for two optics with small losses $\kappa$. 

\bibliographystyle{unsrt85}
\bibliography{bham-ifolab}

\end{document}